\documentclass[twocolumn]{aastex631}

\usepackage{amsmath,amssymb}
\usepackage{graphicx}
\usepackage{booktabs}
\usepackage{xcolor}
\usepackage{hyperref}

\graphicspath{{figures/}{figures/gallery/}{results_comparison/}{results_identifiability/}{results_rar/}{results_mcmc_refined/}{results_mcmc_refined/posterior_predictive/}{results_mcmc_refined/corner_plots/}{results_mcmc_refined/trace_plots/}{results_supplements/}}
\newcommand{\vca}{%
  \ifmmode
    \mathrm{vca}%
  \else
    \textit{VCA }%
  \fi
}
\newcommand{\vinf}{v_\infty}
\newcommand{\rzero}{r_0}
\newcommand{\gbar}{g_\mathrm{bar}}
\newcommand{\gobs}{g_\mathrm{obs}}
\newcommand{\chisq}{\chi^2}
\shorttitle{A velocity-coupled radial-acceleration ansatz for SPARC rotation curves}
\shortauthors{Nalin Dhiman} 

\begin{document}

\title{A Velocity--Coupled Radial Acceleration Ansatz for Disk-Galaxy Rotation Curves:\\ Fits to SPARC, Bayesian Inference, and Parameter Identifiability}

\correspondingauthor{Nalin Dhiman}
\email{d24008@students.iitmandi.ac.in}

\author{Nalin Dhiman}
\affiliation{School of Computing and Electrical Engineering\\ Indian Institute of Technology, Mandi, India}

\vspace{12pt}

\begin{abstract}
Observed rotation curves of disk galaxies remain a sharp empirical probe of the relationship between baryons and dynamics.
We study a minimal, explicitly \emph{phenomenological} alternative to standard halo parameterizations: an additional inward \emph{radial} acceleration proportional to the local \emph{tangential} speed, $a_{\vca}(r)=\gamma(r)\,v(r)$, with a saturating coupling $\gamma(r)=\vinf/(r+\rzero)$.
Combining this ansatz with the circular-motion condition yields a quadratic equation for $v(r)$ with a closed-form physical branch.
We fit this ``velocity-coupled acceleration'' (\vca) model to $N_\mathrm{gal}=171$ rotation curves from the SPARC sample using the published baryonic decompositions (gas, disk, bulge), and we compare to two commonly used two-parameter halo models (NFW and Burkert) using an identical optimization pipeline and error model.
For a fiducial systematic error floor $\sigma_0=5~\mathrm{km\,s^{-1}}$, the \vca model is typically competitive with an NFW halo and performs comparably (though not uniformly better) than a Burkert halo in information-criterion comparisons.
We further perform MCMC inference for \vca parameters, quantify posterior predictive coverage, and show that parameter posteriors exhibit a strong $\vinf$--$\rzero$ degeneracy for many galaxies; only $47/171$ galaxies yield well-identified posteriors under simple width-based criteria. We also perform a simple radial holdout cross-validation (outer 30\% of radii) and find predictive RMSE comparable to NFW and Burkert under this protocol.
Finally, we evaluate how \vca predictions populate the radial acceleration relation (RAR) and find that it reproduces the overall locus with scatter comparable to NFW and somewhat larger than Burkert for this pipeline.
We emphasize that \vca is a compact kinematic fit form rather than a complete dynamical theory: it introduces a velocity-dependent force that is not derived from a conservative potential, and its physical origin remains an open question.
\end{abstract}


\section{Introduction}
Rotation curves of disk galaxies are traditionally modeled by combining the gravitational contributions of luminous matter with either (i) a dark matter halo or (ii) a modification to the inferred gravitational law or inertial response.
The SPARC compilation \citep{Lelli2016} provides high-quality rotation curves with homogeneous Spitzer photometry and baryonic mass models, and it played a central role in establishing the tight empirical radial acceleration relation (RAR; \citealt{McGaugh2016}).

The literature contains many successful parameterizations of rotation curves, including cuspy halos motivated by $\Lambda$CDM simulations (e.g., NFW; \citealt{Navarro1997}), empirically cored halos (e.g., Burkert; \citealt{Burkert1995}), and semi-empirical forms such as the universal rotation curve \citep{Persic1996}.
At the same time, phenomenological regularities like the RAR suggest that, at the level of galaxy kinematics, the ``missing'' acceleration correlates tightly with the baryonic one \citep{McGaugh2016}.

This work explores a deliberately modest question:
\emph{Can one write a compact two-parameter ansatz, distinct from a density-profile halo, that remains self-consistent at the level of circular dynamics and fits SPARC rotation curves competitively?}
We investigate an inward radial acceleration term proportional to the local tangential speed,
\begin{equation}
a_{\vca}(r)=\gamma(r)\,v(r),
\end{equation}
with a simple saturating coupling
\begin{equation}
\gamma(r)=\frac{\vinf}{r+\rzero}.
\end{equation}
This ``velocity--coupled acceleration'' (\vca) form was originally motivated as an attempt to capture, in an effective way, the empirical need for additional radial support in outer disks without committing to a specific halo density profile.
We emphasize upfront that this is a \emph{phenomenological} force law: it is not derived from an action, and it introduces velocity-dependence that generally precludes a global conservative potential.
The aim is to quantify how far such a minimal kinematic rule can go, what it predicts for scaling relations and the RAR, and where it fails.

Our contribution is primarily methodological and empirical.
We (i) implement \vca in a way that is algebraically self-consistent (solving the implicit dynamics exactly), (ii) fit it to SPARC using an apples-to-apples pipeline also applied to NFW and Burkert halos, (iii) evaluate sensitivity to an assumed systematic error floor, (iv) perform MCMC inference to expose degeneracies and identifiability, and (v) connect the fitted curves to the RAR.
We keep the physical claims conservative, and we treat the model mainly as a compact descriptive tool and a possible hint of underlying regularities.

\section{Model}
\subsection{Baryonic contribution}
For each galaxy we adopt the SPARC baryonic decomposition \citep{Lelli2016} in terms of rotation-curve components for gas, stellar disk, and (when present) bulge:
\begin{equation}
v_\mathrm{bar}^2(r)=v_\mathrm{gas}^2(r)+\Upsilon_\mathrm{disk}\,v_\mathrm{disk}^2(r)+\Upsilon_\mathrm{bulge}\,v_\mathrm{bulge}^2(r),
\label{eq:vbar}
\end{equation}
where $\Upsilon_\mathrm{disk}$ and $\Upsilon_\mathrm{bulge}$ are stellar mass-to-light ratios.
In the fiducial analysis we fix $\Upsilon_\mathrm{disk}=0.5$ and $\Upsilon_\mathrm{bulge}=0.7$ (in solar units), which is a commonly adopted choice in SPARC-based analyses \citep{Lelli2016,McGaugh2016}.
Allowing $\Upsilon$ to vary introduces additional degeneracies and is deferred to future work.

The baryonic radial acceleration is then
\begin{equation}
\gbar(r)=\frac{v_\mathrm{bar}^2(r)}{r}.
\end{equation}

\subsection{Velocity--coupled radial acceleration (\vca)}
We postulate an additional inward radial acceleration of the form
\begin{equation}
a_{\vca}(r)=\gamma(r)\,v(r), \qquad
\gamma(r)=\frac{\vinf}{r+\rzero},
\label{eq:gamma}
\end{equation}
where $v(r)$ is the \emph{total} circular speed.
The parameter $\vinf$ has dimensions of velocity and sets an asymptotic scale; $\rzero$ is a length scale controlling how rapidly the coupling saturates with radius.
The choice in Eq.~\eqref{eq:gamma} is the simplest two-parameter form that (i) has correct units, (ii) avoids a divergent coupling at $r\to 0$, and (iii) yields a finite outer amplitude.

For circular motion, the radial force balance is
\begin{equation}
\frac{v^2(r)}{r} = \gbar(r) + a_{\vca}(r).
\label{eq:circular}
\end{equation}
Multiplying Eq.~\eqref{eq:circular} by $r$ and using Eq.~\eqref{eq:gamma} gives an implicit quadratic equation for $v(r)$:
\begin{equation}
v^2(r) = v_\mathrm{bar}^2(r) + A(r)\,v(r),
\qquad
A(r)\equiv \frac{\vinf\,r}{r+\rzero}.
\label{eq:quadratic}
\end{equation}
Solving Eq.~\eqref{eq:quadratic} yields
\begin{equation}
v(r)=\frac{1}{2}\left[A(r)+\sqrt{A^2(r)+4\,v_\mathrm{bar}^2(r)}\right],
\label{eq:vsolution}
\end{equation}
where we choose the positive branch to ensure $v(r)\ge 0$.
Equation~\eqref{eq:vsolution} is a central practical feature of \vca: although the acceleration depends on the speed, the circular speed is still available in closed form.

\subsubsection{Limiting behavior}
Two limits are useful for interpretation.
At small radii $r\ll \rzero$, $A(r)\approx (\vinf/\rzero)\,r$ and the extra term is perturbative, so $v(r)\approx v_\mathrm{bar}(r)$ when baryons dominate.
At large radii $r\gg \rzero$, $A(r)\to \vinf$ and Eq.~\eqref{eq:quadratic} becomes $v^2-\vinf v - v_\mathrm{bar}^2\simeq 0$, implying $v\to \vinf$ when the baryonic term declines.
Thus \vca naturally tends to asymptotically flat rotation curves without explicitly adding a halo density profile.

\subsubsection{Effective ``halo'' interpretation on circular orbits}
Although \vca is not equivalent to adding an independent mass component, on a \emph{circular} orbit one can define an effective extra radial acceleration
\begin{equation}
g_{\vca}(r)\equiv \gobs(r)-\gbar(r)=\gamma(r)\,v(r)=\frac{\vinf}{r+\rzero}\,v(r),
\end{equation}
and (purely for intuition) a corresponding effective enclosed mass,
\begin{equation}
M_{\vca}(<r)\equiv \frac{r^2\,g_{\vca}(r)}{G}.
\end{equation}
If $r\gg \rzero$ and the curve is approximately flat, $v(r)\simeq \vinf$, then $g_{\vca}\simeq \vinf^2/r$ so that $M_{\vca}(<r)\propto r$ and the implied effective density scales as $\rho_{\vca}\propto r^{-2}$, as in an isothermal sphere.
If instead $r\ll \rzero$ (so $\gamma\simeq \vinf/\rzero$ is nearly constant) and the inner rotation is approximately solid body, $v(r)\simeq \Omega r$, then $g_{\vca}\propto r$ and $M_{\vca}(<r)\propto r^3$, corresponding to an approximately constant effective density in the very inner region.

These correspondences are intended only to build intuition about how \vca can mimic cored or isothermal-like behavior \emph{in circular equilibrium}.
Because the acceleration depends on the instantaneous speed, \vca does not define a unique time-independent potential or density profile for general motions.

\subsubsection{Energy and ``drag''}
Because the \vca acceleration is \emph{radial} while the circular velocity is \emph{tangential}, the instantaneous power on a strictly circular orbit is $P=m\,\mathbf{a}\cdot\mathbf{v}=0$.
In that limited sense, Eq.~\eqref{eq:gamma} should not be interpreted as a literal dissipative drag force.
However, a velocity-dependent force that is not derivable from a scalar potential is nonstandard and may exchange energy on noncircular or time-dependent trajectories.
A full dynamical completion (e.g., an underlying field theory or an effective description in which the force is always perpendicular to $\mathbf{v}$, as in Lorentz-type forces) is beyond the scope of this phenomenological study.
We therefore treat \vca as a kinematic ansatz intended only for circular rotation-curve fitting.

\subsubsection{An ``effective'' extra contribution}
Equation~\eqref{eq:quadratic} can be rearranged as
\begin{equation}
v^2(r)=v_\mathrm{bar}^2(r)+v_{\vca}^2(r),
\qquad
v_{\vca}^2(r)\equiv A(r)\,v(r),
\label{eq:veff}
\end{equation}
which defines an \emph{implicit} effective contribution $v_{\vca}(r)$.
This decomposition is useful for visualization (e.g., ``model anatomy'' plots), but one should remember that $v_{\vca}$ is not an independent additive component: it depends on the total $v$ via Eq.~\ref{eq:vsolution}.

\section{Comparison models: NFW and Burkert halos}
To benchmark \vca against standard halo parameterizations, we fit two widely used two-parameter dark matter profiles with the same data model and optimization settings.

\subsection{NFW halo}
The NFW density profile \citep{Navarro1997} is
\begin{equation}
\rho_\mathrm{NFW}(r)=\frac{\rho_s}{(r/r_s)\,(1+r/r_s)^2},
\end{equation}
with scale density $\rho_s$ and scale radius $r_s$.
The enclosed mass is
\begin{equation}
M_\mathrm{NFW}(r)=4\pi\rho_s r_s^3\left[\ln(1+x)-\frac{x}{1+x}\right],
\qquad x\equiv \frac{r}{r_s},
\end{equation}
and the halo circular speed is $v_\mathrm{NFW}^2(r)=G M_\mathrm{NFW}(r)/r$.
The total model is $v^2=v_\mathrm{bar}^2+v_\mathrm{NFW}^2$.

\subsection{Burkert halo}
The Burkert profile \citep{Burkert1995} is an empirical cored halo,
\begin{equation}
\rho_\mathrm{B}(r)=\frac{\rho_0\,r_b^3}{(r+r_b)\,(r^2+r_b^2)},
\end{equation}
with central density $\rho_0$ and core radius $r_b$.
The enclosed mass has a closed form,
\begin{equation}
\begin{split}
M_\mathrm{B}(r)&=\pi \rho_0 r_b^3\big[\ln(1+x^2)\\
&\quad +2\ln(1+x)-2\arctan(x)\big],\\
x&\equiv \frac{r}{r_b},
\end{split}
\end{equation}
and $v_\mathrm{B}^2(r)=G M_\mathrm{B}(r)/r$.
Again, $v^2=v_\mathrm{bar}^2+v_\mathrm{B}^2$.

\section{Data and inference pipeline}
\subsection{SPARC sample and preprocessing}
We use the SPARC rotation-curve files \citep{Lelli2016} containing $(r_i, v_i, \sigma_i)$ and the baryonic components needed to compute $v_\mathrm{bar}(r)$.
We apply the same quality cuts to all models: galaxies must have at least five usable radial points after basic cleaning.
This yields $N_\mathrm{gal}=171$ galaxies in the present analysis.
Example input files are included in the accompanying data folder.

\subsection{Error model and likelihood}
Following common practice in rotation-curve fitting, we inflate the quoted uncertainties with a systematic error floor,
\begin{equation}
\sigma_{\mathrm{eff},i}^2 = \sigma_i^2 + \sigma_0^2,
\label{eq:sigmaeff}
\end{equation}
where $\sigma_0$ accounts for noncircular motions and other unmodeled systematics.
We explore $\sigma_0\in\{0,3,5,8\}~\mathrm{km\,s^{-1}}$ and adopt $\sigma_0=5~\mathrm{km\,s^{-1}}$ as a fiducial choice for most figures.

Assuming independent Gaussian errors in $v$, the log-likelihood is
\begin{equation}
\ln \mathcal{L} = -\frac{1}{2}\sum_i\left[\frac{(v_i-v_\mathrm{mod}(r_i))^2}{\sigma_{\mathrm{eff},i}^2}+\ln\left(2\pi\sigma_{\mathrm{eff},i}^2\right)\right].
\end{equation}
For model \emph{comparison} we report information criteria using the usual $\chisq$ approximation, which differs only by constants when the same $\sigma_{\mathrm{eff},i}$ are used across models for a given galaxy:
\medmuskip=3mu
\thinmuskip=2mu
\begin{equation}
\chisq=\sum_i\frac{(v_i-v_\mathrm{mod}(r_i))^2}{\sigma_{\mathrm{eff},i}^2},\;
\mathrm{AIC}=\chisq+2k,\;
\mathrm{BIC}=\chisq+k\ln N,
\end{equation}
\medmuskip=4mu plus 2mu minus 4mu 
\thinmuskip=3mu
with $k$ parameters and $N$ data points \citep{Akaike1974,Schwarz1978}.

\subsection{Deterministic fits and bounds}
We obtain best-fit parameters for each galaxy and model via bounded nonlinear least squares (SciPy; \citealt{Virtanen2020}), using the \emph{same} optimizer configuration and broad bounds for all models.
For \vca, the fitted parameters are $(\log_{10}\vinf,\log_{10}\rzero)$ with bounds $\log_{10}\vinf\in(-1,4)$ and $\log_{10}\rzero\in(-2,3)$, corresponding to $\vinf\in(0.1,10^4)\,\mathrm{km\,s^{-1}}$ and $\rzero\in(0.01,10^3)\,\mathrm{kpc}$.
Analogously broad bounds are used for NFW and Burkert parameters.
We explicitly track ``bound hits'' as a diagnostic of non-identifiability.

\subsection{Bayesian inference, posterior predictive checks, and identifiability}
For the \vca model we additionally perform MCMC inference for $\sigma_0=5~\mathrm{km\,s^{-1}}$ using the affine-invariant ensemble sampler \citep{GoodmanWeare2010,ForemanMackey2013}.
We adopt uniform priors over the same log-bounds as the deterministic fits.
For each galaxy, we monitor the integrated autocorrelation time $\tau$ for both parameters and run chains long enough to satisfy a conservative $N_\mathrm{step}/\tau \gtrsim 50$ criterion.
Across the sample we find median $\tau$ values of $\tau(\log\vinf)\approx 26.0$ and $\tau(\log\rzero)\approx 26.3$ steps, with median acceptance fraction $\approx 0.55$.

We compute posterior predictive intervals by drawing parameter samples from the posterior, evaluating $v_\mathrm{mod}(r)$, and adding Gaussian measurement noise with $\sigma_\mathrm{eff}$.
Coverage is quantified by the fraction of observed points falling inside the nominal $68\%$ and $95\%$ predictive intervals.

Finally, we quantify parameter \emph{identifiability} using posterior widths in log-space and a simple lever-arm statistic.
We define three tiers:
(A) well-constrained if the $68\%$ widths in both $\log\vinf$ and $\log\rzero$ are $<0.5$ dex and $R_\mathrm{max}/\rzero>2$,
(B) moderately constrained if only the width criterion holds,
(C) unconstrained otherwise.
This classification is meant as a transparent diagnostic rather than a deep statement about the physics.

\subsection{RAR diagnostics}
We compute observed and baryonic radial accelerations as
\begin{equation}
\gobs(r)=\frac{v_\mathrm{obs}^2(r)}{r},\qquad
\gbar(r)=\frac{v_\mathrm{bar}^2(r)}{r},
\end{equation}
and we compare $\gobs(\gbar)$ against model predictions $g_\mathrm{mod}(r)=v_\mathrm{mod}^2(r)/r$.
We summarize RAR performance using the RMS scatter (in dex) of $\log_{10}g_\mathrm{mod}-\log_{10}\gobs$ over all radial points.

\section{Results}
\subsection{Fit quality and information criteria}
\begin{figure*}
\centering
\includegraphics[width=0.32\textwidth]{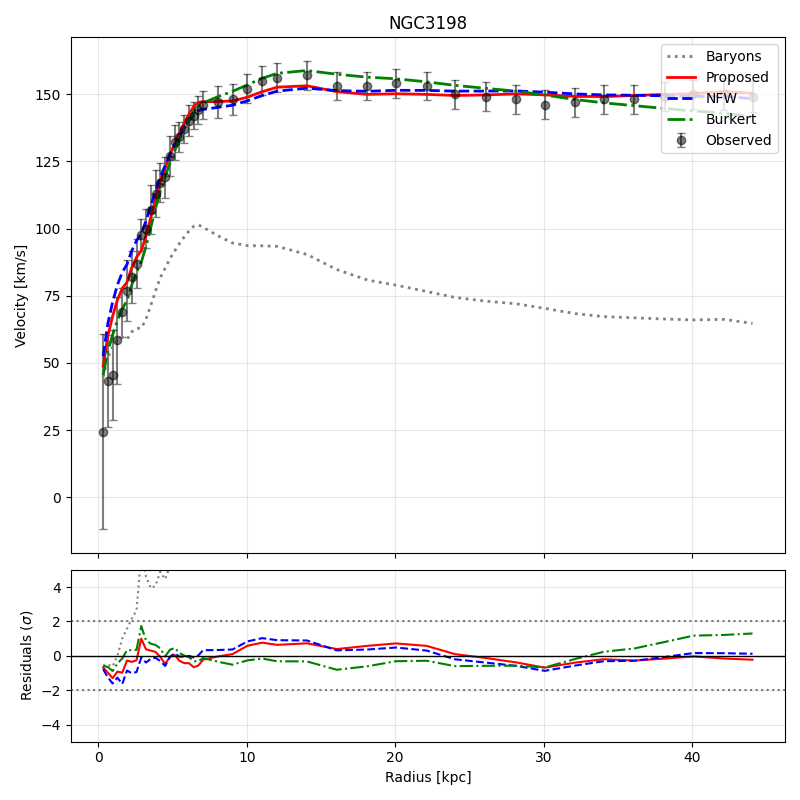}
\includegraphics[width=0.32\textwidth]{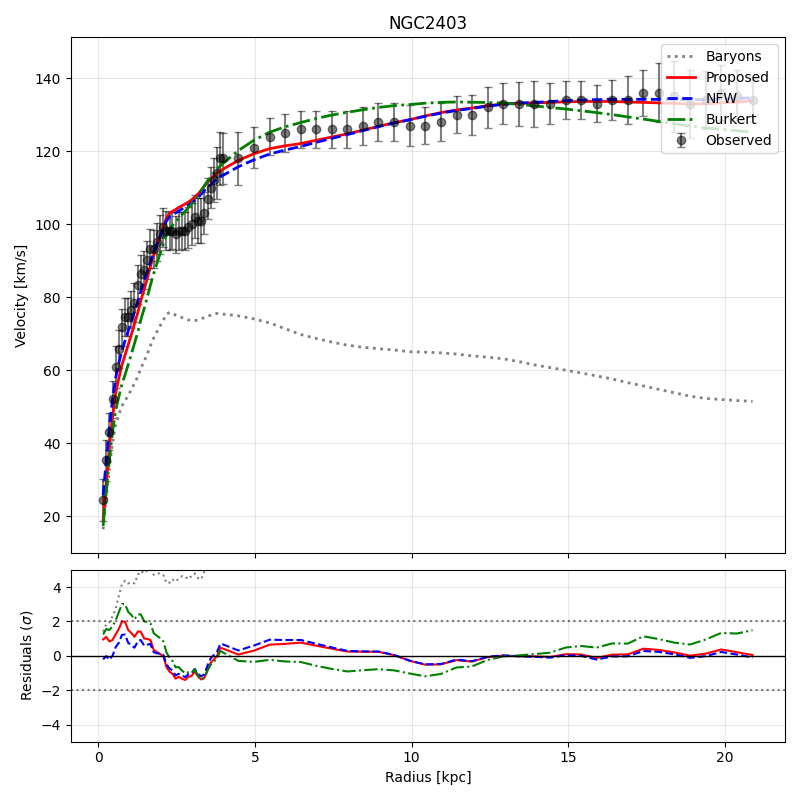}
\includegraphics[width=0.32\textwidth]{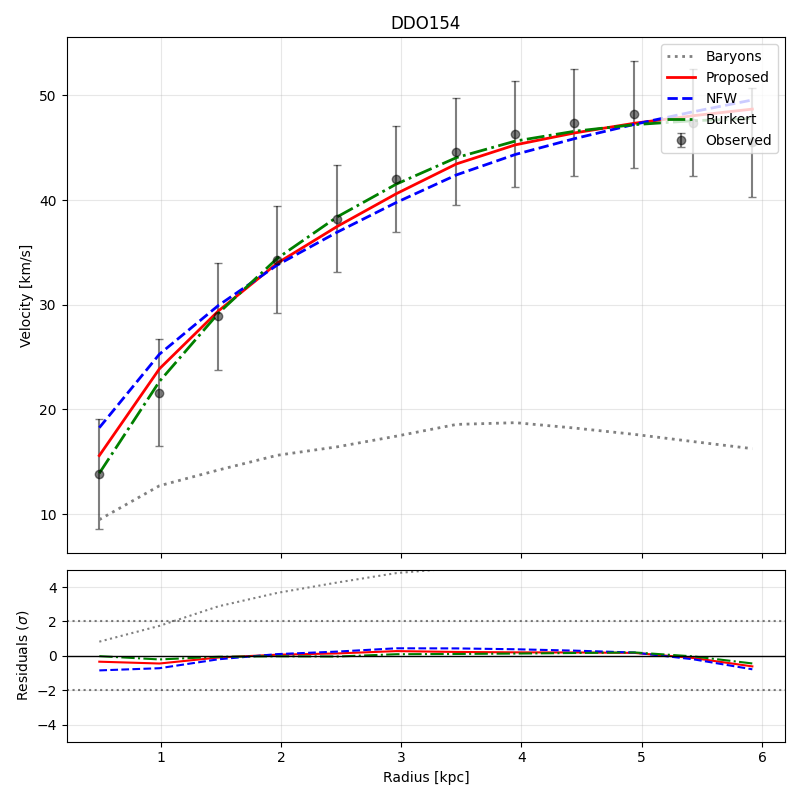}
\caption{Representative SPARC rotation curves with best-fit models and residuals (in units of $\sigma_\mathrm{eff}$; Eq.~\ref{eq:sigmaeff}). Shown are a high-quality extended spiral (NGC3198), an intermediate case (NGC2403), and a dwarf (DDO154). The \vca model (red) is competitive with NFW (blue) and Burkert (green) in many cases, though performance varies galaxy-to-galaxy.}
\label{fig:example_fits}
\end{figure*}

Figure~\ref{fig:example_fits} shows representative rotation-curve fits and standardized residuals for three galaxies spanning different morphologies and velocity scales.
All three two-parameter models (\vca, NFW, Burkert) fit markedly better than baryons-only, as expected.

Across the full sample and adopting $\sigma_0=5~\mathrm{km\,s^{-1}}$, the median reduced chi-square values are
$\tilde{\chisq}_\nu\approx 29.9$ (baryons-only),
0.39 (\vca),
0.53 (NFW),
and 0.29 (Burkert).
In terms of AIC, \vca tends to slightly outperform NFW (median $\Delta\mathrm{AIC}_{\mathrm{NFW}-\vca}\approx 0.30$), while Burkert is slightly favored over \vca in the median (median $\Delta\mathrm{AIC}_{\mathrm{Bur}-\vca}\approx -0.60$).

\begin{figure*}
\centering
\includegraphics[width=0.85\textwidth]{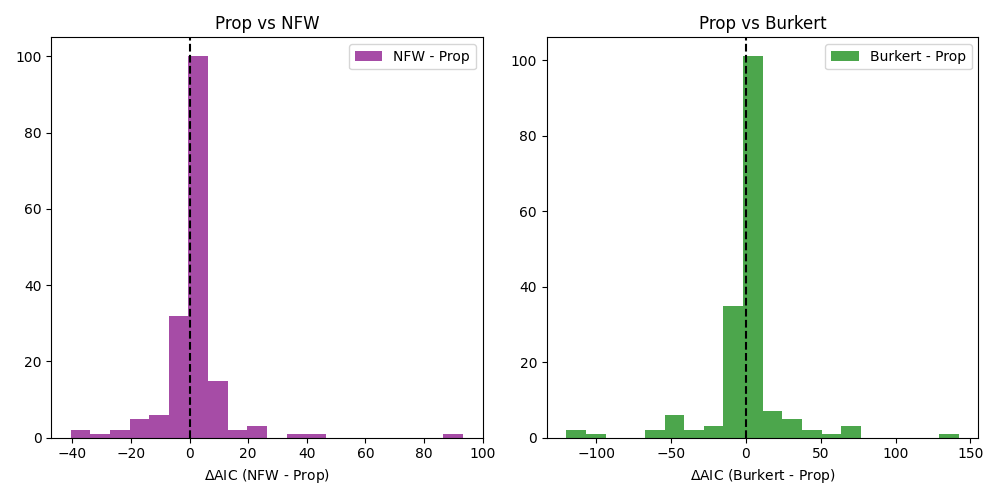}
\caption{AIC comparison at $\sigma_0=5~\mathrm{km\,s^{-1}}$. Left: $\Delta\mathrm{AIC}=\mathrm{AIC}_\mathrm{NFW}-\mathrm{AIC}_\vca$ (positive favors \vca). Right: $\Delta\mathrm{AIC}=\mathrm{AIC}_\mathrm{Bur}-\mathrm{AIC}_\vca$ (positive favors \vca).}
\label{fig:aicdist}
\end{figure*}

Figure~\ref{fig:aicdist} shows the distributions of $\Delta\mathrm{AIC}$ at $\sigma_0=5~\mathrm{km\,s^{-1}}$.
The tails demonstrate that no two-parameter model wins universally, and that a meaningful subset of galaxies prefer one parameterization over another.

We summarize sensitivity to the systematic error floor $\sigma_0$ in Table~\ref{tab:sensitivity} (included from the pipeline output).
The qualitative conclusions are stable across reasonable $\sigma_0$ values: \vca is competitive with NFW in AIC for a majority of galaxies, whereas Burkert is more often preferred over \vca but with a large ``tie'' fraction when using a $\Delta\mathrm{AIC}\le 2$ equivalence threshold.

\begin{table}
\centering
\begin{tabular}{l|cccc|cc}
\hline
$\sigma_0$ [km/s] & \multicolumn{4}{c|}{Median $\chi^2_\nu$} & \multicolumn{2}{c}{VCA AIC Win \% vs} \\
 & Bar & VCA & NFW & Bur & NFW & Bur \\
\hline
0 & 85.65 & 1.02 & 1.28 & 0.78 & 59.6\% & 33.9\% \\
3 & 49.52 & 0.57 & 0.80 & 0.45 & 59.6\% & 32.7\% \\
5 & 29.86 & 0.39 & 0.53 & 0.29 & 61.4\% & 32.2\% \\
8 & 15.96 & 0.22 & 0.29 & 0.16 & 63.2\% & 32.2\% \\
\hline
\end{tabular}
\caption{Median reduced $\chi^2$ and AIC model preference fractions as a function of the systematic error floor $\sigma_0$.}
\label{tab:sensitivity}
\end{table}

\subsection{Radial holdout cross-validation}
\label{subsec:cv}

Information criteria such as AIC quantify in-sample descriptive adequacy under a chosen likelihood,
but phenomenological models can still overfit in ways that do not generalize across radius.
To assess out-of-sample performance, we perform a simple radial holdout cross-validation within each galaxy:
we fit each model using the inner 70\% of measured radii (ordered by $r$) and evaluate predictive error on the outer 30\%.
We report the test-set root-mean-square error (RMSE) in km~s$^{-1}$ on the withheld outer points.
Only galaxies with a sufficient number of points in both splits are included ($N_{\rm gal}=143$ in this run).

\begin{figure}
  \centering
  \includegraphics[width=\linewidth]{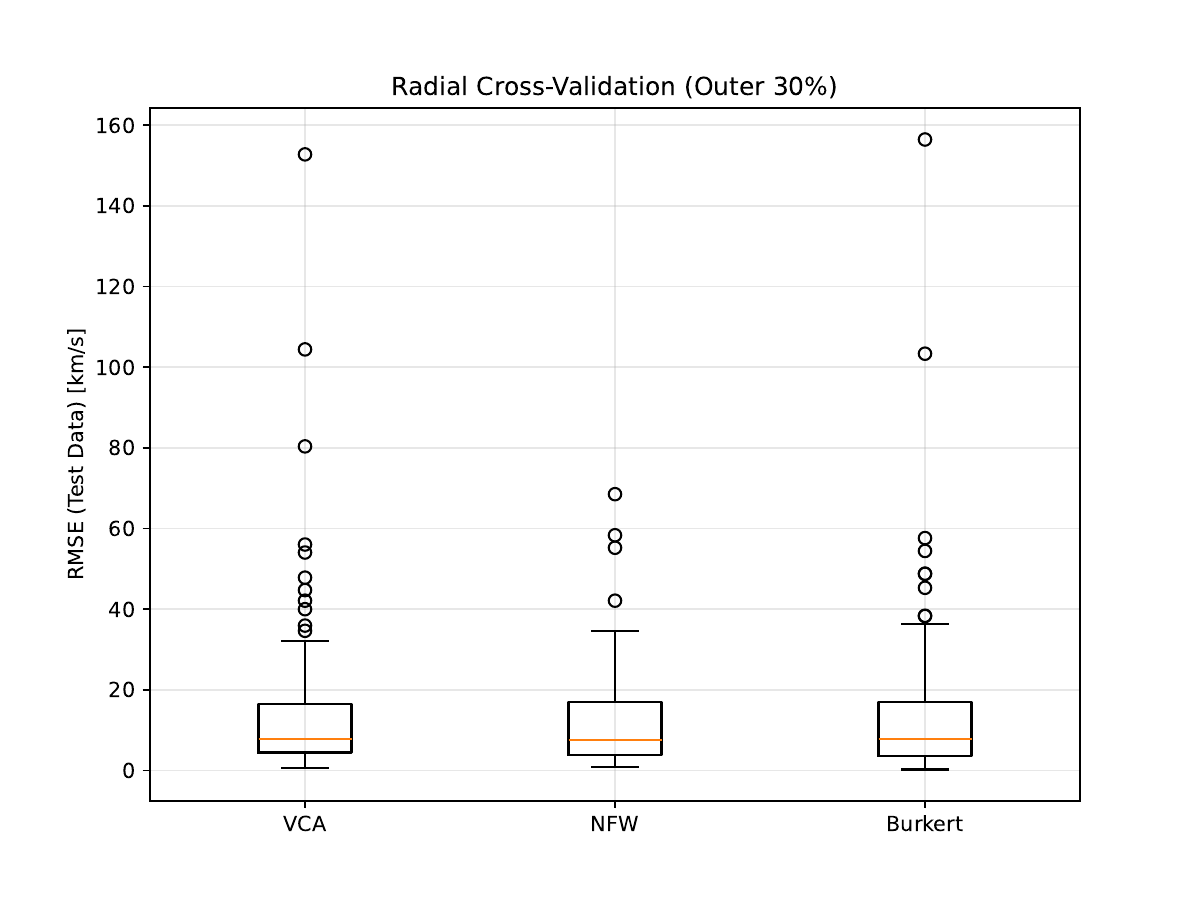}
  \caption{Radial cross-validation using the outer 30\% of radii as a test set.
  The RMSE distributions for VCA, NFW, and Burkert are broadly similar, indicating comparable out-of-sample
  performance for this radial holdout protocol.}
  \label{fig:cv_rmse}
\end{figure}

Figure~\ref{fig:cv_rmse} summarizes the distribution of test RMSE for VCA, NFW, and Burkert.
The three models exhibit comparable median predictive accuracy on the outer radii.
Paired signed-rank tests on per-galaxy RMSE differences do not indicate a systematic advantage of any model in this metric.

\begin{table}
\centering
\caption{Radial cross-validation summary (outer 30\% test set; $N_{\rm gal}=143$).}
\label{tab:cv_summary}
\begin{tabular}{lccc}
\toprule
Model & Median RMSE & 16--84\% RMSE & Win fraction \\
 & (km s$^{-1}$) & (km s$^{-1}$) & (lowest RMSE) \\
\midrule
VCA     & 7.86 & 3.08--22.82 & 0.259 \\
NFW     & 7.64 & 3.01--21.99 & 0.315 \\
Burkert & 7.90 & 2.31--26.16 & 0.413 \\
\bottomrule
\end{tabular}
\end{table}

\subsection{Model anatomy}
\begin{figure}
\centering
\includegraphics[width=\linewidth]{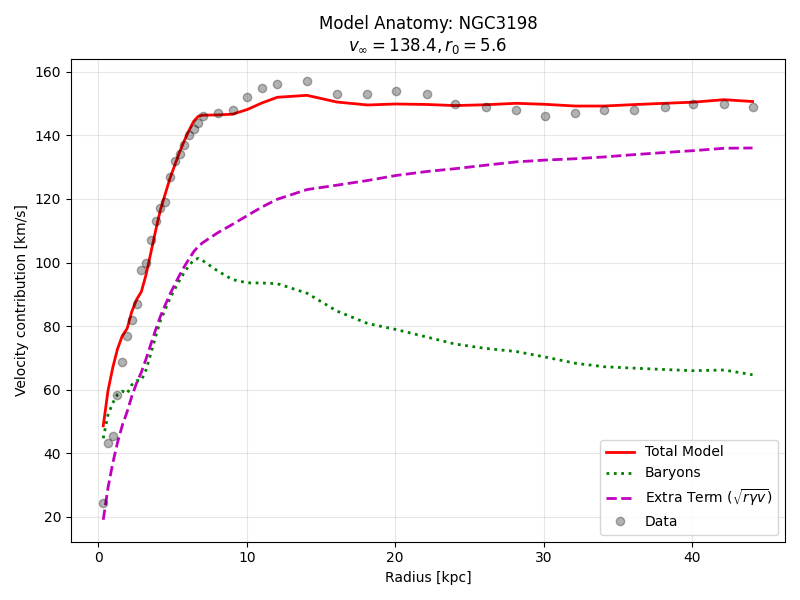}
\caption{Model anatomy for NGC3198. The implicit effective contribution $v_{\vca}(r)$ (Eq.~\ref{eq:veff}) rises and saturates, producing a flat outer rotation curve when combined with the baryonic component.}
\label{fig:anatomy}
\end{figure}

To illustrate how the implicit extra term behaves, Figure~\ref{fig:anatomy} shows a ``model anatomy'' decomposition (Eq.~\ref{eq:veff}) for NGC3198.
The baryonic contribution peaks in the inner disk and declines, while the effective \vca term grows and saturates, yielding an approximately flat outer curve.
This behavior is qualitatively similar to a transition from baryon-dominated inner dynamics to halo-dominated outer dynamics, but it is achieved through a velocity-dependent coupling rather than an explicit mass density profile.

\subsection{Bayesian posteriors, degeneracy, and identifiability}
\begin{figure}
\centering
\includegraphics[width=\linewidth]{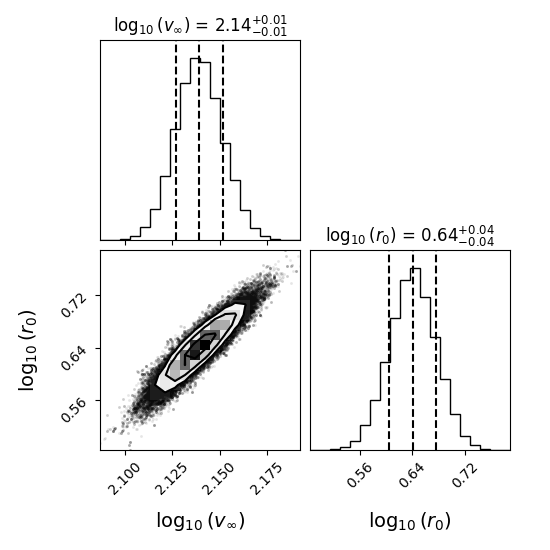}
\caption{Example \vca posterior (NGC2403). The $\log\vinf$--$\log\rzero$ degeneracy is visible but still yields relatively tight marginal constraints.}
\label{fig:corner_example}
\end{figure}

MCMC posteriors typically show a strong correlation between $\log\vinf$ and $\log\rzero$, consistent with the fact that rotation curves constrain combinations of parameters through $A(r)=\vinf r/(r+\rzero)$.
Figure~\ref{fig:corner_example} shows an example corner plot for a well-constrained case (NGC2403).
In contrast, many galaxies exhibit broad, banana-shaped posteriors and/or bound hits, indicating that the data do not separately identify $\vinf$ and $\rzero$ under our assumptions.

\begin{figure}
\centering
\includegraphics[width=\linewidth]{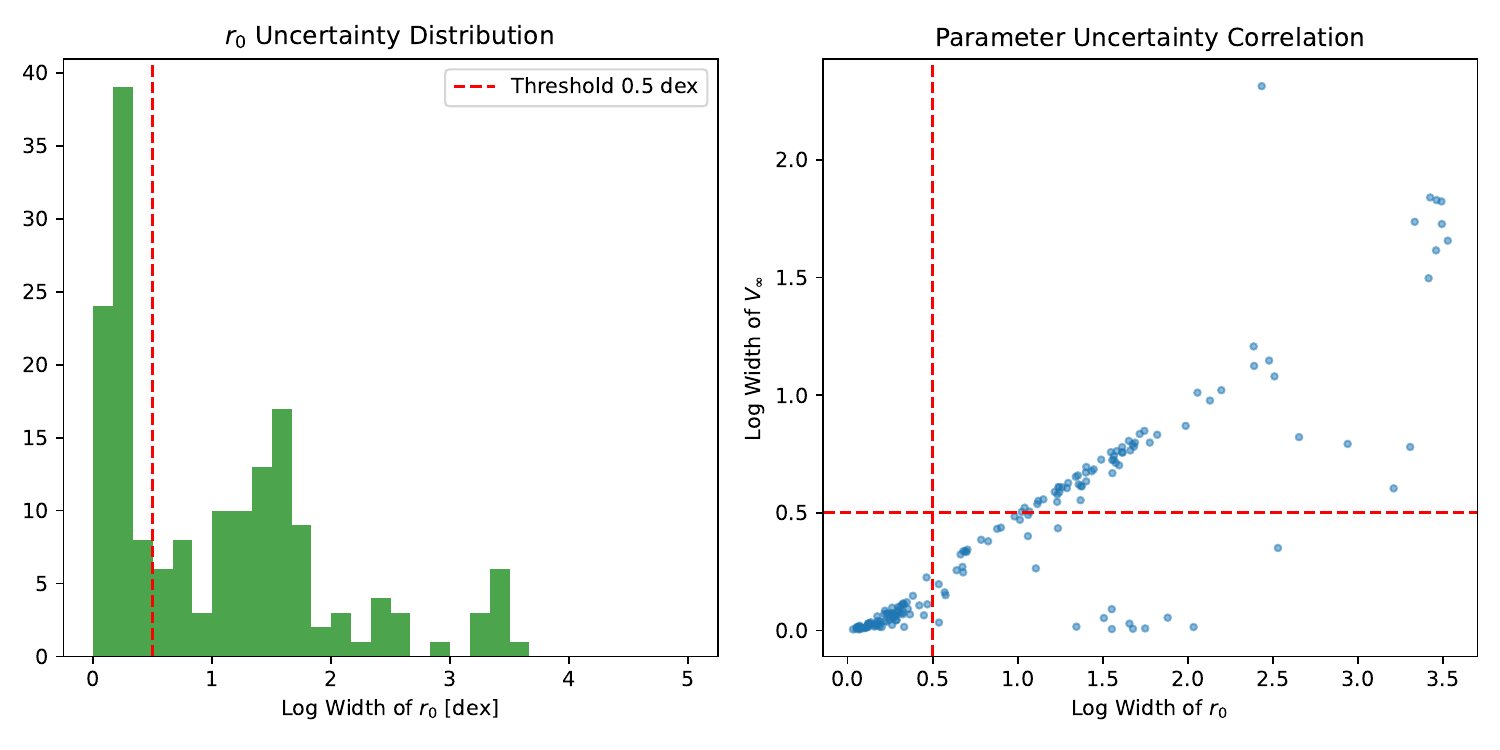}
\caption{Identifiability diagnostics for \vca posteriors. Left: distribution of $\log$-widths for $\rzero$ (68\% credible interval); right: correlation between $\log$-widths of $\vinf$ and $\rzero$. The dashed lines mark the 0.5-dex threshold used in Table~\ref{tab:identifiability}.}
\label{fig:ident_hist}
\end{figure}

We quantify identifiability across the sample using the tiered classification described above.
Table~\ref{tab:identifiability} summarizes the results: only $47/171$ galaxies meet the Tier A criterion.
Figure~\ref{fig:ident_hist} shows the distribution of posterior widths and their correlation, emphasizing that broad posteriors are common even when chains converge.

\begin{table}
\centering
\begin{tabular}{lcc}
\hline
Tier & Definition & Count (\%) \\
\hline
A (Well-Constrained) & $\Delta < 0.3$ dex, $R_{max}/r_0 > 2$ & 47 (27.5\%) \\
B (Moderately) & $\Delta < 0.5$ dex & 23 (13.5\%) \\
C (Unconstrained) & Loose or Bound Hit & 101 (59.1\%) \\
\hline
Total & & 171 \\
\hline
\end{tabular}
\caption{Classification of galaxy parameter identifiability.}
\label{tab:identifiability}
\end{table}

\begin{figure}
\centering
\includegraphics[width=\linewidth]{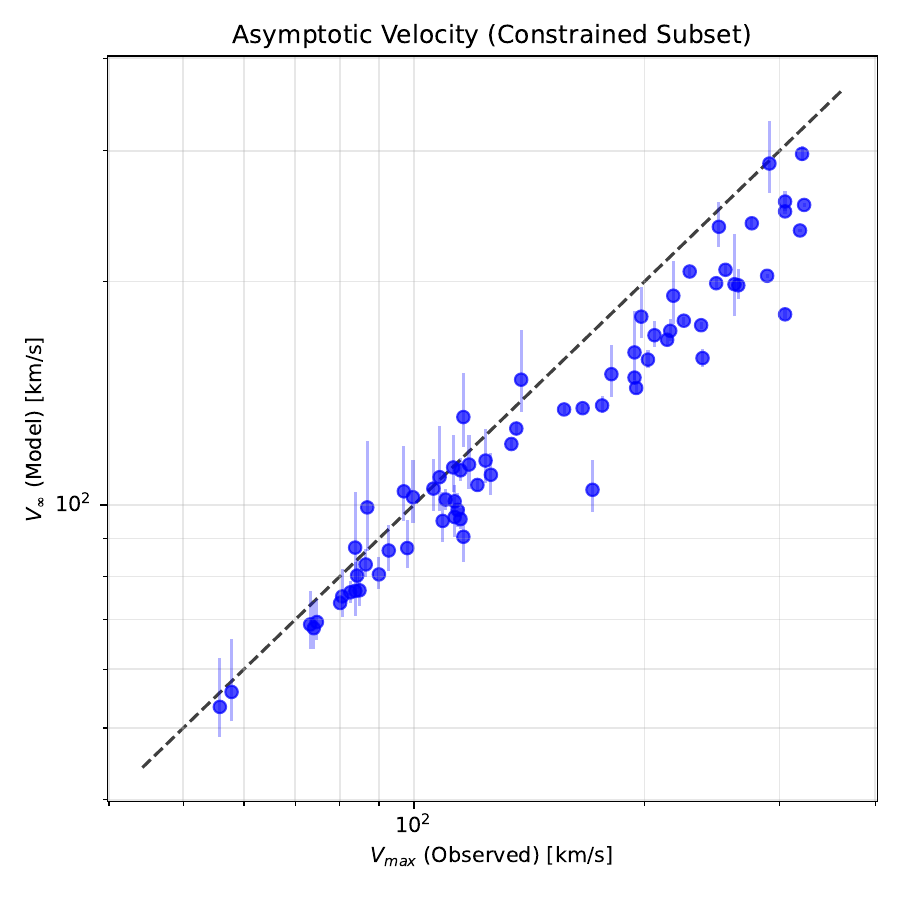}
\caption{Asymptotic parameter $\vinf$ versus observed maximum speed $V_\mathrm{max}$ for the constrained subset. The dashed line indicates equality.}
\label{fig:vinf_vmax}
\end{figure}

For the constrained subset, $\vinf$ is strongly correlated with the observed maximum speed $V_\mathrm{max}$ (Figure~\ref{fig:vinf_vmax}), supporting the interpretation of $\vinf$ as an ``outer'' velocity scale.
However, because $\rzero$ is often poorly constrained, a more robust derived quantity is the effective amplitude $A(R_\mathrm{max})=\vinf R_\mathrm{max}/(R_\mathrm{max}+\rzero)$, which is closely tied to the modeled speed at the outermost measured radius.

\subsection{Posterior predictive checks and uncertainty calibration}
\begin{figure}
\centering
\includegraphics[width=\linewidth]{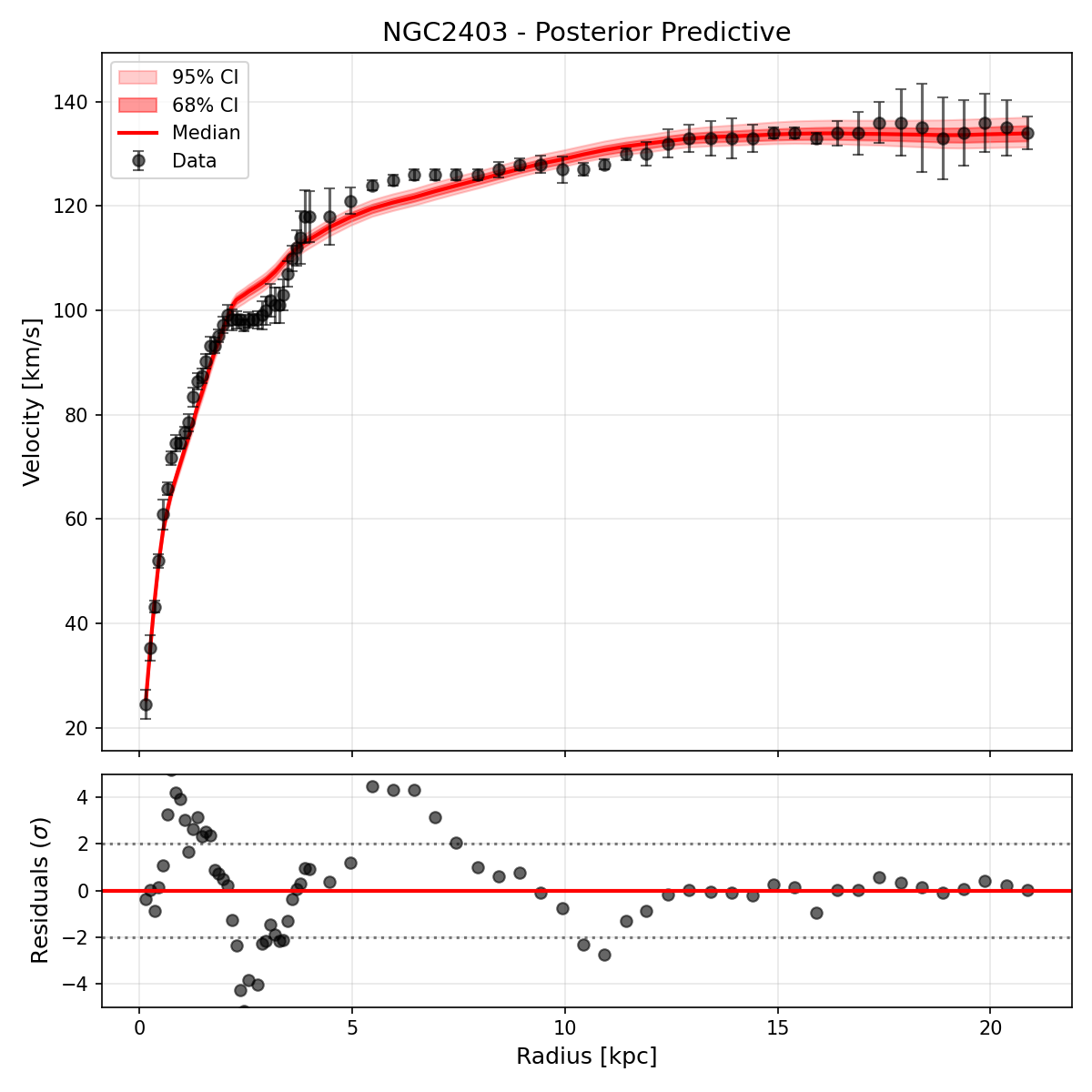}
\caption{Posterior predictive check for NGC2403 (68\% and 95\% credible intervals). Residuals are shown in units of $\sigma_\mathrm{eff}$.}
\label{fig:ppc_example}
\end{figure}

A critical question for any Bayesian fit is whether its predictive intervals are calibrated.
Figure~\ref{fig:ppc_example} shows a posterior predictive summary for NGC2403.
Aggregating over all galaxies and radii, we find global predictive coverage of 77.3\% for nominal $68\%$ intervals and 93.6\% for nominal $95\%$ intervals.
The $95\%$ coverage is close to nominal, while the $68\%$ intervals tend to be conservative (slightly over-covering), consistent with an error model that includes a systematic floor.

\subsection{Radial acceleration relation}
\begin{figure}
\centering
\includegraphics[width=\linewidth]{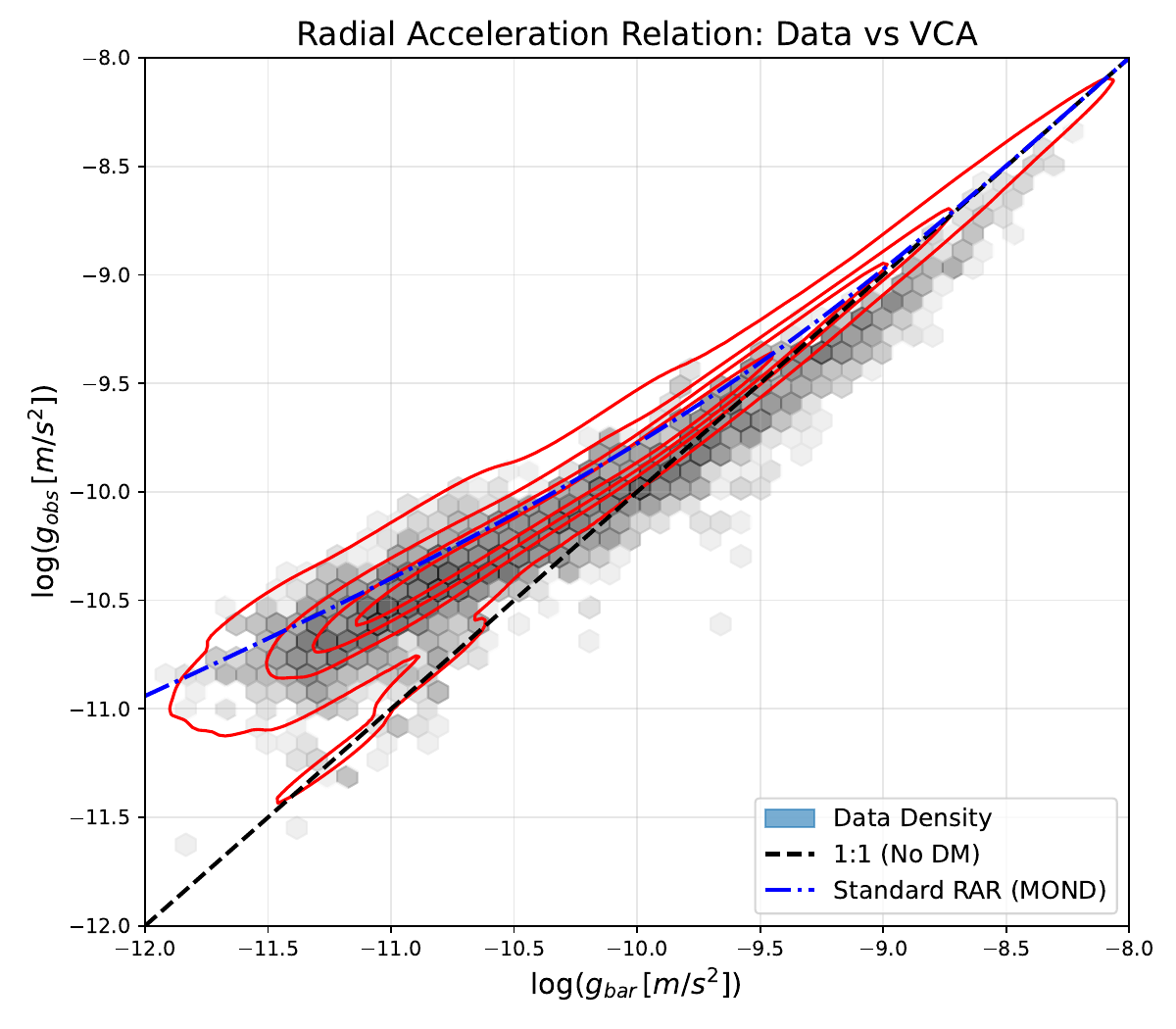}
\caption{RAR diagnostic. The gray density shows the data distribution in $(\gbar,\gobs)$ space. The dashed line marks $\gobs=\gbar$ (no discrepancy) and the dash-dotted curve shows a standard MOND-like RAR for reference \citep{Milgrom1983,McGaugh2016}. Red contours indicate the \vca model population.}
\label{fig:rar}
\end{figure}

Figure~\ref{fig:rar} places \vca predictions in the $(\gbar,\gobs)$ plane.
Qualitatively, the model reproduces the main bend away from the $1{:}1$ line.
Quantitatively, Table~\ref{tab:rar_scatter} reports the RMS scatter (dex) of model accelerations relative to observed accelerations across all radial points.
Within this pipeline, \vca yields scatter comparable to NFW and somewhat larger than Burkert.

\begin{table}
\centering
\begin{tabular}{lc}
\hline
Model & RMS Scatter (dex) \\
\hline
VCA (Proposed) & 0.196 \\
NFW & 0.199 \\
Burkert & 0.177 \\
\hline
\end{tabular}
\caption{Global scatter of the Radial Acceleration Relation residuals relative to observed acceleration.}
\label{tab:rar_scatter}
\end{table}

\section{Discussion: interpretation, limitations, and reviewer-proofing}
\subsection{Why consider \vca if Burkert often fits better?}
A fair criticism is that the Burkert halo an empirical cored density profile often outperforms \vca in AIC for this sample, while using the same number of free parameters.
We do \emph{not} claim that \vca replaces Burkert as a practical fitting function.

The motivation for \vca is different.
A Burkert halo is a parametric \emph{mass profile}; it fits by assigning an additional, effectively ``invisible'' mass distribution.
By contrast, \vca is a parametric \emph{force law} (albeit phenomenological) that couples the extra radial acceleration to the \emph{observed kinematics} through $a_{\vca}\propto v$.
Even when fit quality is similar, these are conceptually distinct hypotheses.
If one ultimately seeks dynamical mechanisms that connect baryons and kinematics (as suggested empirically by the RAR), testing such couplings may be informative.
Our results show that \vca is at least competitive with NFW in many galaxies and reproduces the gross structure of the RAR, but they also show that Burkert remains a stronger empirical baseline in this framework.

\subsection{The ``drag'' pitfall and how we avoid it}
The historical ``drag'' intuition (a force proportional to speed) is problematic if interpreted as a tangential dissipative force.
Here the acceleration is \emph{radial}, so it does no work on a perfectly circular orbit.
Nonetheless, because it depends on the speed, it is not generated by a time-independent scalar potential and therefore is not automatically consistent with energy and angular momentum conservation for general trajectories.
This is the price of phenomenological flexibility.
We therefore present \vca explicitly as a kinematic rule for circular motion, and we refrain from claims about orbit families, stability, or cosmological consistency.

If one wishes to pursue a physical origin, plausible directions include (i) Lorentz-like effective forces that are always perpendicular to $\mathbf{v}$ (so they do no work), (ii) modified-inertia frameworks in which the relation between force and acceleration depends on the state of motion, or (iii) effective descriptions of complex baryon--dark-sector interactions.
These avenues would require an explicit covariant formulation and additional observational tests (e.g., lensing, vertical kinematics, dynamical friction), which we do not attempt here.

\subsection{Parameter degeneracy and what is actually constrained}
The strong $\vinf$--$\rzero$ degeneracy means that interpreting $\vinf$ and $\rzero$ individually as physical scales is often unjustified.
In practice, rotation curves constrain the function $A(r)=\vinf r/(r+\rzero)$ over the observed radial range.
When the data do not extend far beyond $\rzero$, the saturation is weakly sampled and $(\vinf,\rzero)$ trade off along near-degenerate directions.
This is why derived quantities tied to the observed radial lever arm (e.g., $A(R_\mathrm{max})$ or $v(R_\mathrm{max})$) are typically more robust.
A useful extension would be to re-parameterize \vca in terms of such derived quantities, which may improve identifiability without changing the model family.

\subsection{Limitations and opportunities for improvement}
Several simplifying assumptions limit the interpretation of our results:
\begin{enumerate}
\item \textbf{Fixed stellar mass-to-light ratios.} Allowing $\Upsilon_\mathrm{disk}$ and $\Upsilon_\mathrm{bulge}$ to vary with informative priors would better reflect astrophysical uncertainty but would increase degeneracy.
\item \textbf{Distance and inclination uncertainties.} SPARC provides distance estimates and quality flags \citep{Lelli2016}. A hierarchical treatment could propagate those uncertainties into parameter posteriors.
\item \textbf{Limited cross-validation.}
We include a simple radial holdout cross-validation (Section~\ref{subsec:cv}) that tests extrapolation from inner to outer radii.
This is only one out-of-sample protocol; more stringent tests (e.g., leave-one-galaxy-out on scaling relations,
or hierarchical Bayesian predictive checks) remain for future work.
\item \textbf{Not a full theory.} Because \vca is not derived from a potential, it is not immediately applicable to lensing or cosmological structure formation.
\end{enumerate}

\section{Conclusions}
We introduced and tested a two-parameter velocity--coupled radial acceleration ansatz (\vca) for disk galaxy rotation curves.
The model is algebraically self-consistent for circular motion and has a closed-form solution (Eq.~\ref{eq:vsolution}).

Using an apples-to-apples fitting pipeline on $N_\mathrm{gal}=171$ SPARC galaxies, we find:
\begin{enumerate}
\item \vca is typically competitive with an NFW halo in fit quality and AIC, and it produces realistic flat outer rotation curves.
\item A Burkert halo remains the stronger empirical baseline in this comparison, but \vca is often comparable within $\Delta\mathrm{AIC}\le 2$.
\item MCMC inference reveals that parameter identifiability is limited for many galaxies; only $47/171$ are well-constrained under simple criteria.
\item In the $(\gbar,\gobs)$ plane, \vca reproduces the gross RAR locus with scatter comparable to NFW and somewhat larger than Burkert.
\item A simple radial holdout cross-validation yields comparable median predictive RMSE among VCA, NFW, and Burkert for the outer radii.

\end{enumerate}
We therefore view \vca as a compact descriptive model and a diagnostic of baryon--kinematics couplings, rather than as a replacement for physical halo models.
A credible next step would be to develop a dynamical completion and to test \vca like couplings against additional observables beyond circular rotation.

\clearpage
\appendix
\section{Dependence of model preference on galaxy properties}
\label{app:daic_props}

\begin{figure*}
  \centering
  \includegraphics[width=0.95\textwidth]{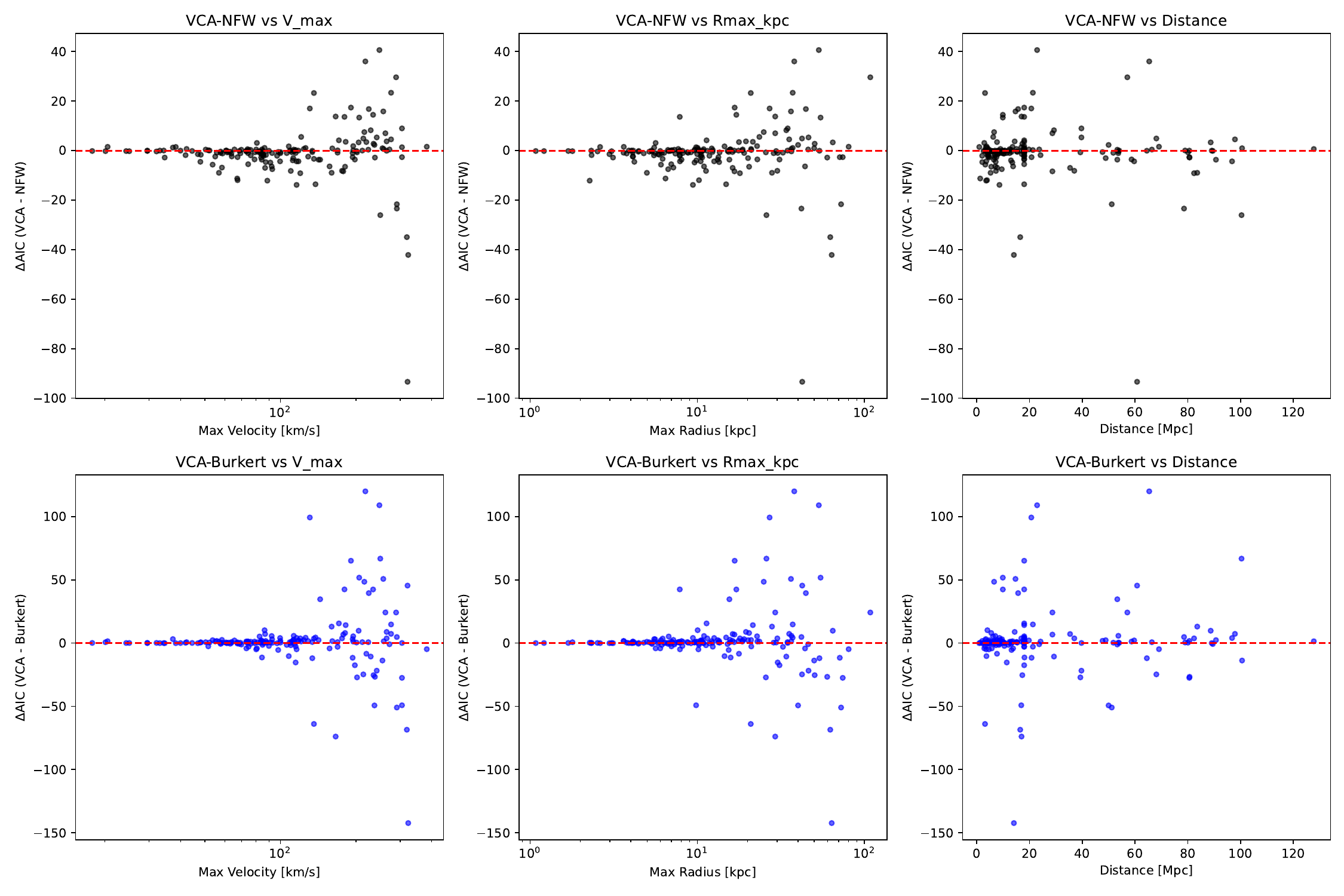}
  \caption{Model-preference diagnostics as a function of galaxy properties.
  Here $\Delta\mathrm{AIC}$ is plotted as \emph{VCA minus comparison model}
  (negative values favor VCA).
  We find no strong monotonic dependence on distance and only weak dependence on $V_{\max}$;
  the dominant effect is increased scatter at large $V_{\max}$ and large $R_{\max}$.}
  \label{fig:daic_props}
\end{figure*}

To assess whether relative model preference is associated with basic observational properties,
we examine $\Delta\mathrm{AIC}$ as a function of $V_{\max}$, the maximum sampled radius $R_{\max}$, and distance.
Figure~\ref{fig:daic_props} shows these comparisons for VCA relative to NFW and Burkert.

\section{Additional figures and diagnostic material}
To keep the main text focused, we include several diagnostic figures here.

\begin{figure*}
\centering
\includegraphics[width=0.85\textwidth]{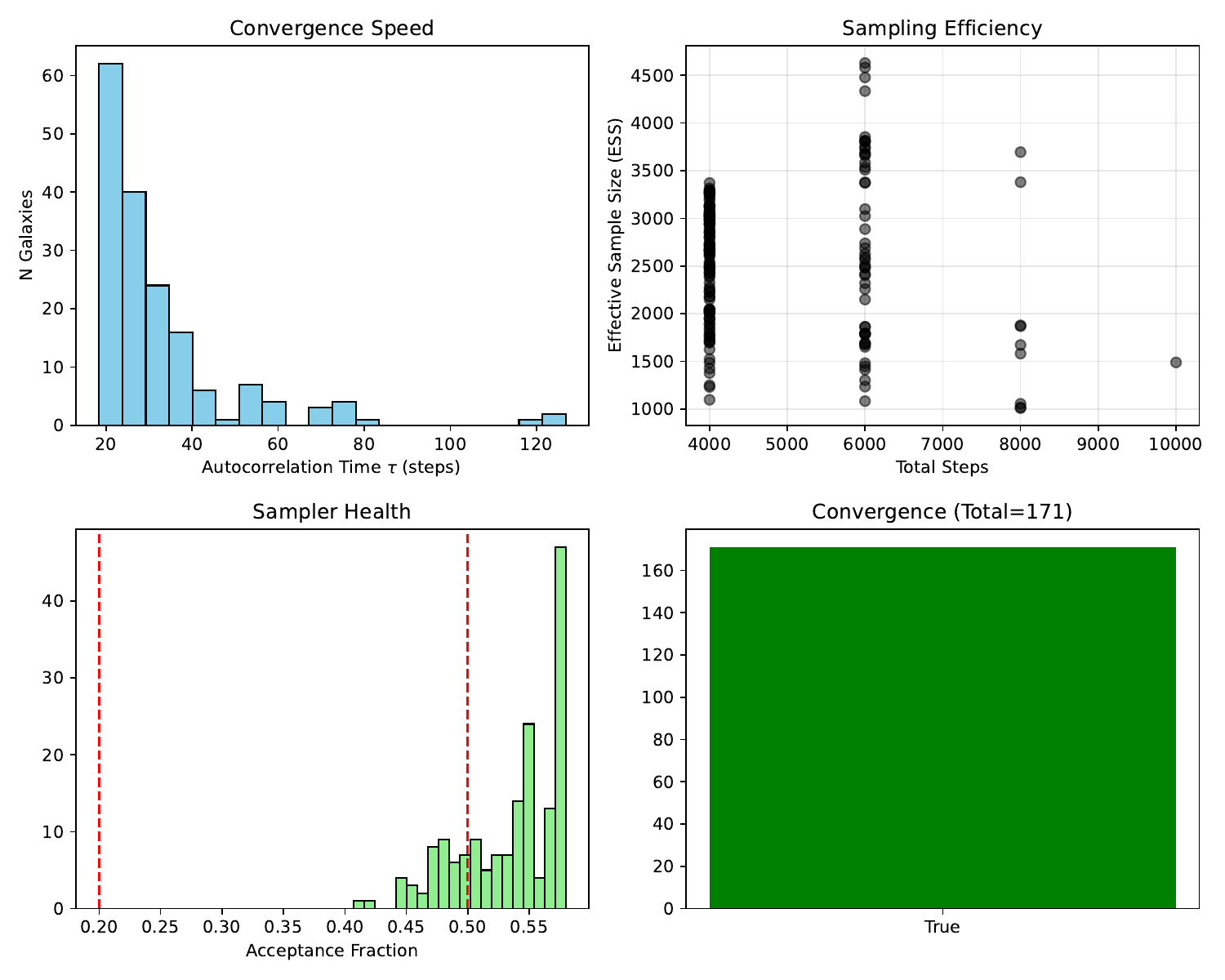}
\caption{MCMC diagnostic summary across the sample: distributions of autocorrelation times, effective sample sizes, and acceptance fractions.}
\label{fig:mcmc_diag}
\end{figure*}

\begin{figure*}
\centering
\includegraphics[width=0.85\textwidth]{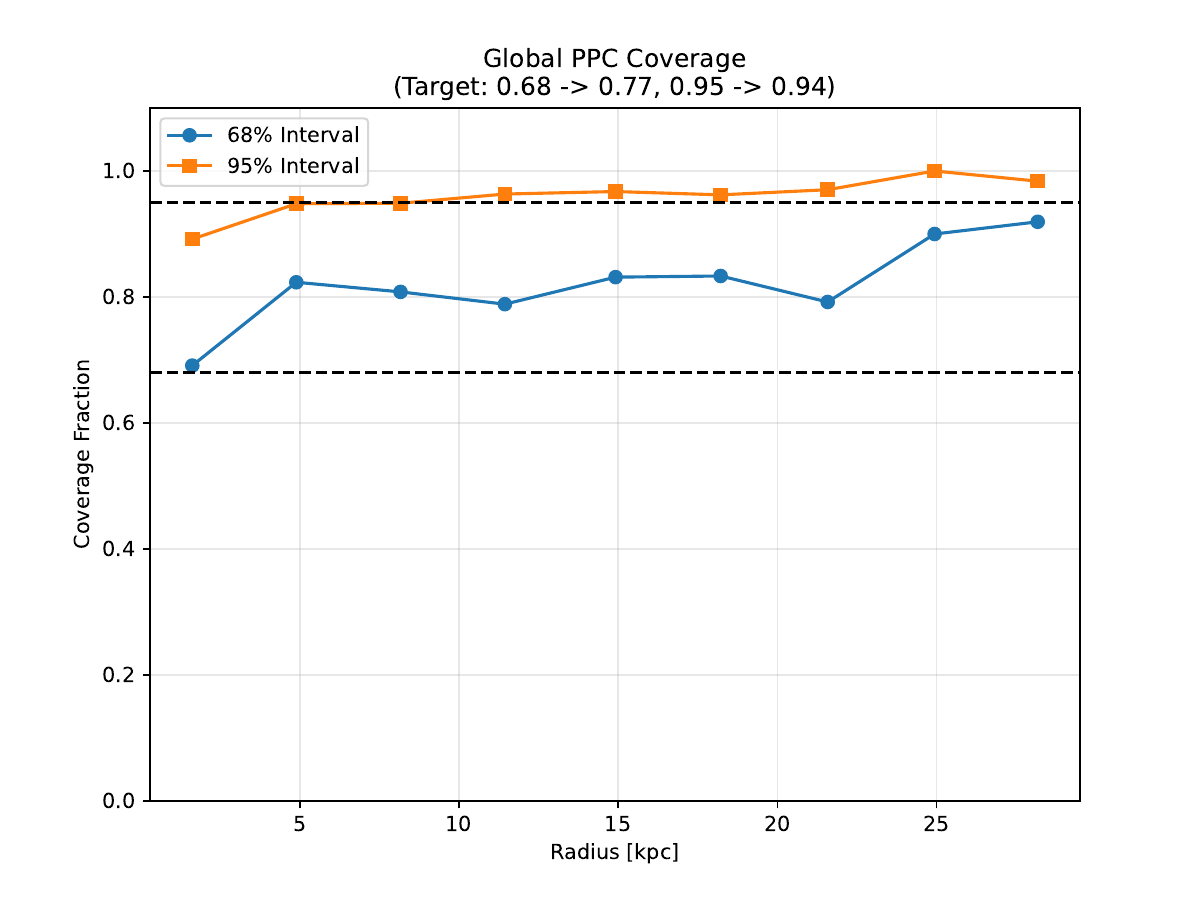}
\caption{Posterior predictive coverage as a function of radius. Nominal 68\% and 95\% predictive intervals are shown; 68\% intervals are somewhat conservative under the adopted error model.}
\label{fig:coverage}
\end{figure*}

\begin{figure*}
\centering
\includegraphics[width=0.85\textwidth]{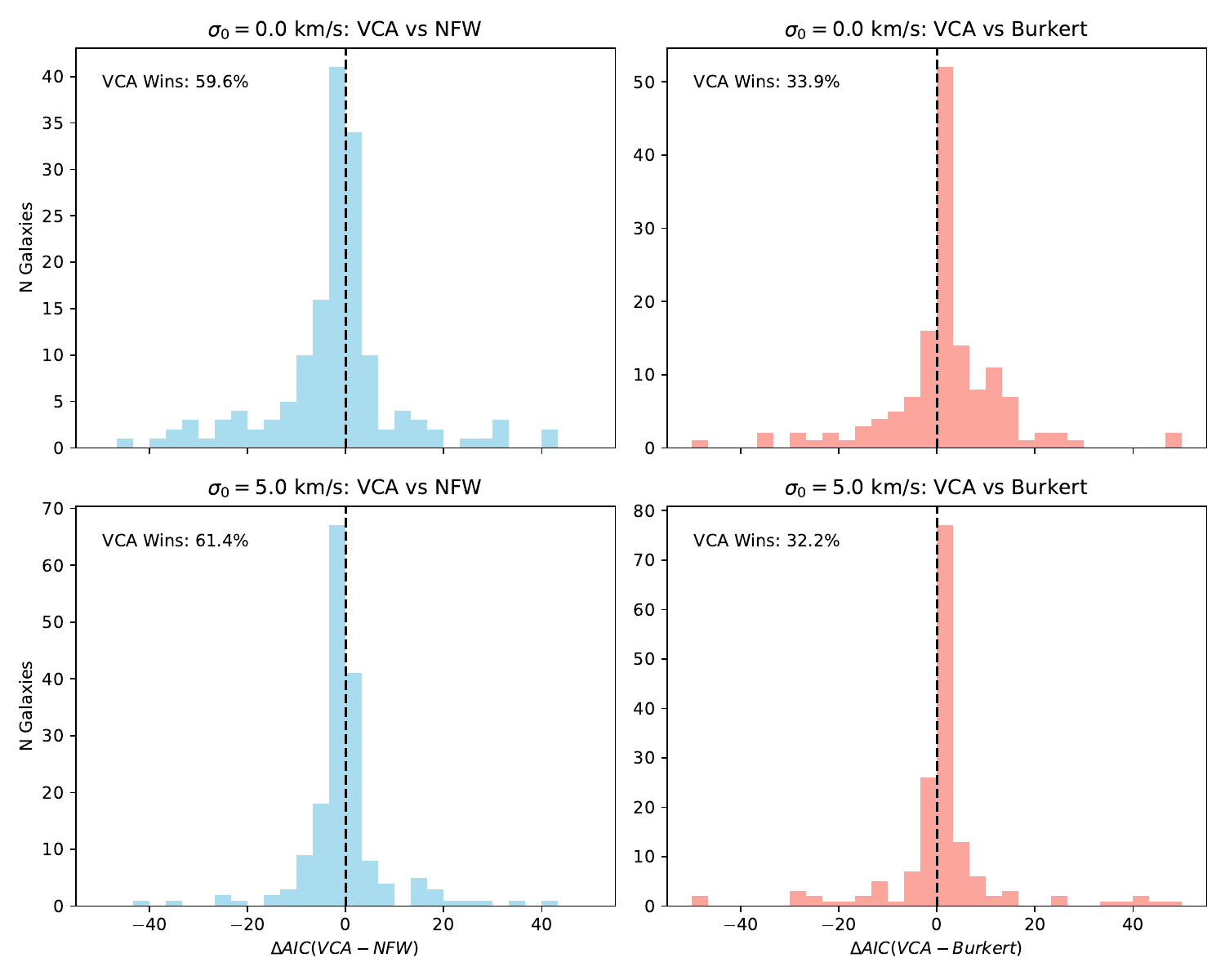}
\caption{Sensitivity of AIC differences to the systematic error floor $\sigma_0$.}
\label{fig:sensitivity_hist}
\end{figure*}

\begin{figure}
\centering
\includegraphics[width=\linewidth]{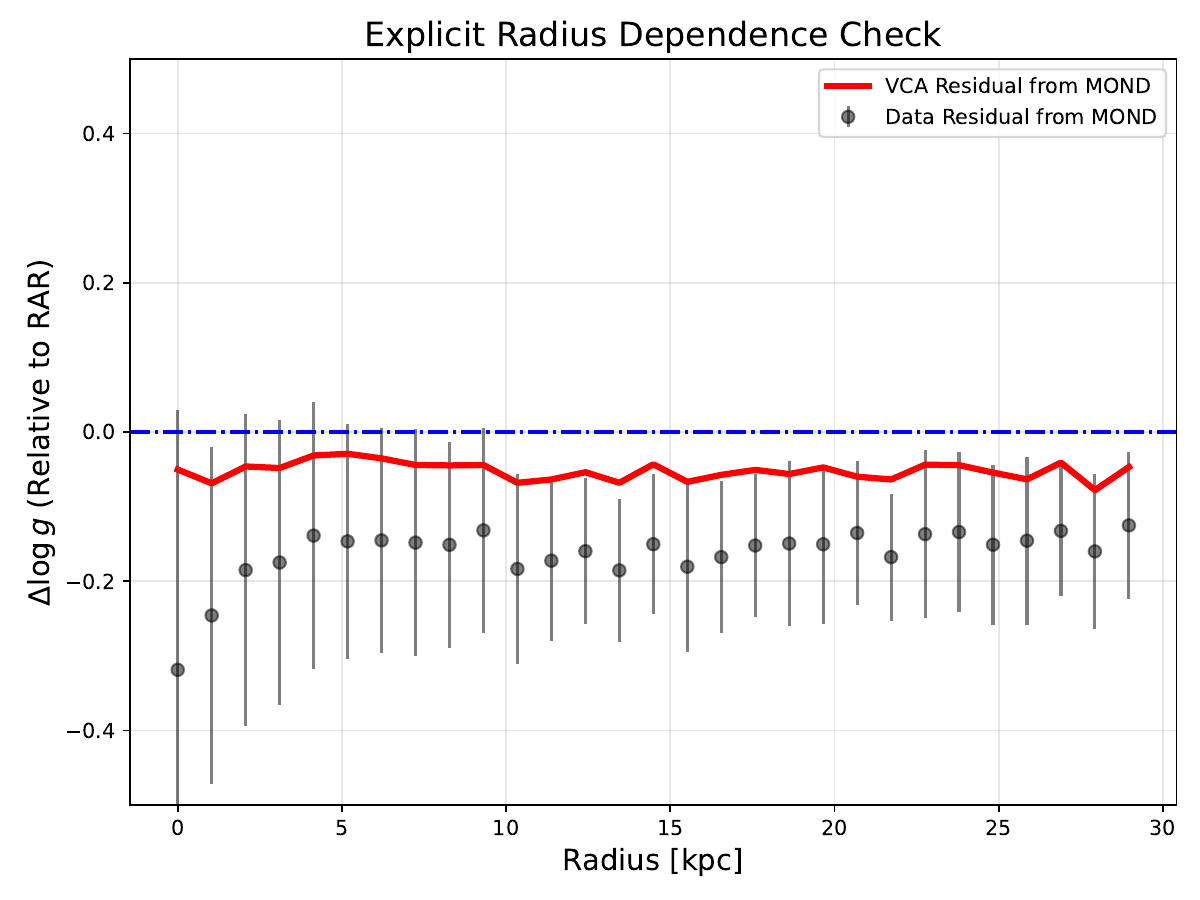}
\caption{RAR residuals as a function of radius (pipeline diagnostic).}
\label{fig:rar_resid_r}
\end{figure}

\begin{figure}
\centering
\includegraphics[width=\linewidth]{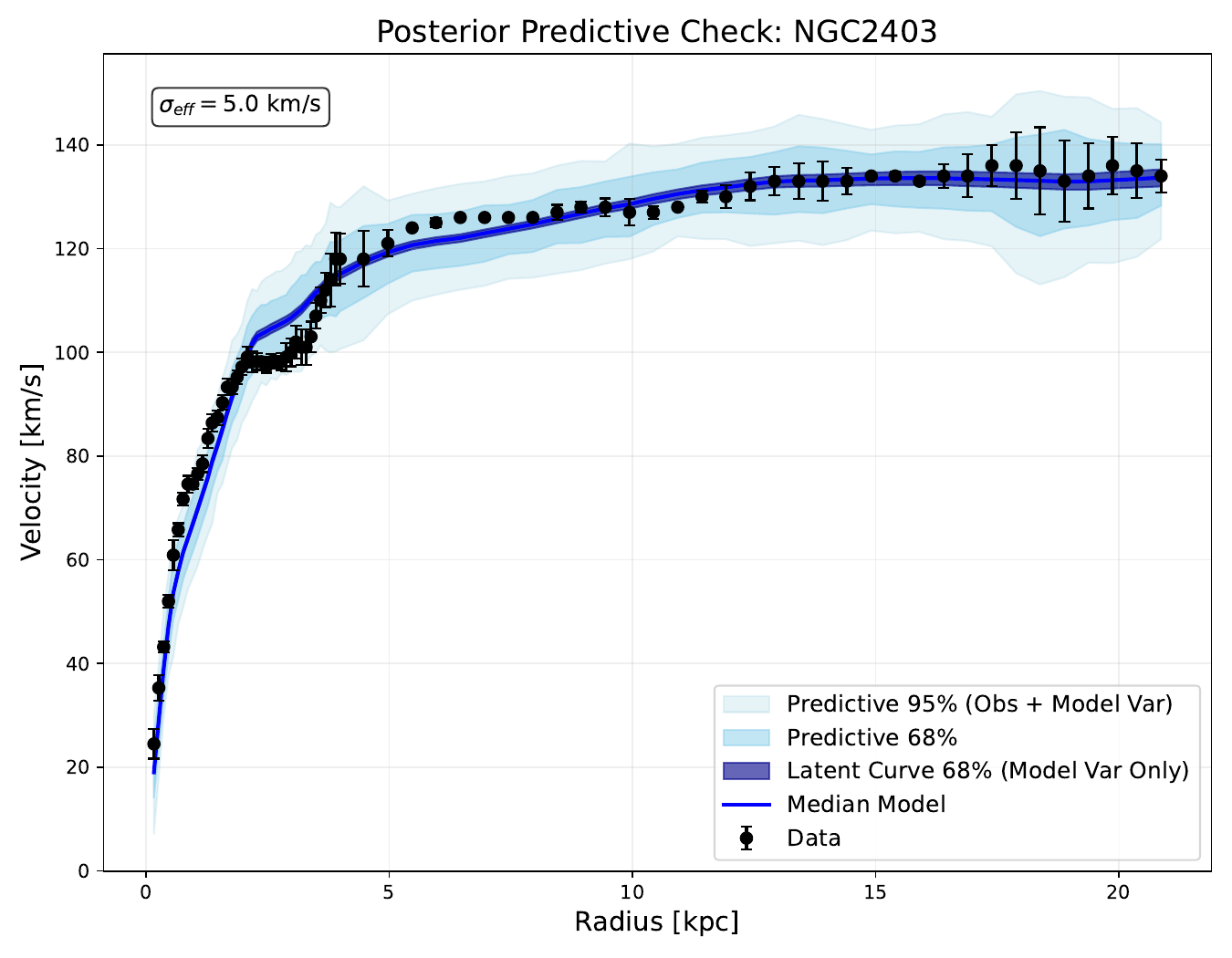}
\caption{Schematic illustrating latent-curve and posterior-predictive uncertainty for NGC2403.}
\label{fig:schematic_ppc}
\end{figure}

\subsection{Supplementary online material}
The full per-galaxy fit gallery (171 panels with residuals) is provided as \texttt{figures/all\_fits\_gallery.pdf}.
A compiled set of all MCMC corner plots is provided as \texttt{results\_mcmc\_refined/all\_corner\_plots.pdf}.
Machine-readable fit results and model-comparison metrics are provided as CSV files in the accompanying repository.\href{https://github.com/nalin-dhiman/VCA_Rotmod_LTG.git}{Github}


\end{document}